\documentclass[superscriptaddress,twocolumn,amsmath,amssymb,groupedaddress]{revtex4-1}
\usepackage{graphicx}
\usepackage{dcolumn}
\usepackage{bm}
\usepackage[dvipsnames]{xcolor}

\begin{document}

\title{Collective Behaviour of Composite Active Particles}

\author{Joshua Eglinton}
\author{Mike I. Smith}
\author{Michael R. Swift}
\affiliation{School of Physics and Astronomy, University of Nottingham, Nottingham NG7 2RD, UK}

\begin{abstract}
We describe simulations of active Brownian particles carried out to
explore how dynamics and clustering are influenced by particle shape.
Our particles are composed of four disks, held together by springs, whose relative size can be varied.
These composite objects can be tuned smoothly from having a predominantly concave
to a convex surface. We show that even two of these composite particles can 
exhibit collective motion which modifies the effective Peclet number. We then investigate how particle geometry can be used to explain the phase behaviour of many such particles.
\end{abstract}

\maketitle

There is currently significant interest in the dynamics of active
matter and the modelling of collective biological motion \cite{Popkin}. Active
particles take their energy from their surroundings to
induce directed motion.
Natural examples of collective motion in active systems include the flocking of birds, 
shoals of fish and clustering in bacterial colonies \cite{Marchetti}.
Within the Physics community, active matter represents a
statistical system driven far from equilibrium which
can can exhibit non-equilibrium ordering and phase
transitions \cite{Jaeger}. However, to date, there is no overarching
framework to describe such phenomena.

Much of the theoretical work in this field has focused on
collections of freely rotating disks or spheres \cite{Fily,Redner1}. 
Under appropriate conditions, such systems can exhibit clustering,
often referred to as motility induced phase separation (MIPS) \cite{Cates}.
This transition results from a competition between the rate of arrival
of particles in a region of space and the ability of particles to rotate and 
leave \cite{Solon1,Takatori}.
In these systems
the addition of aligning interactions is known to enhance clustering \cite{Sese}, whilst attractive interactions can significantly influence the collective 
behaviour \cite{Schwarz,Redner2}. 

These ideas have been explored in more complex systems via experiments and simulations
to investigate asymmetric active particles
including rods \cite{Bar,Nagai,Narayan}, dumbbells \cite{Cugliandolo} and 
composite or continuum systems \cite {Weitz,Menzel,Grossmann}. It is now realised
that particle shape can have a significant influence on the
collective dynamics.

One further feature that active systems have in common is the
strong coupling between the particles' dynamics and their
interaction with boundaries \cite{Bechinger}. It is known that active
systems can drive non-equilibrium interfacial fluctuations \cite{Junot},
exhibit non-ideal pressure variations \cite{Solon2} and show clustering effects 
induced by the presence of a boundary \cite{Deblais}. Intriguingly 
the convex or concave nature
of the boundary can have a pronounced effect on the
clustering that is observed \cite{Kumar}.

In this Article we describe simulations of an active particle
system in which the shape of the particles can be changed continuously.
Specifically, we consider particle shapes which have surfaces that possess
both concave and convex regions.
We investigate how this structure influences
the dynamics of a small number of particles and 
the collective behaviour of many such particles. 

Our basic active particle is constructed out of four disks, as
illustrated in the insets to Fig. 1. The disks are held together by
five springs so that the composite particle, referred
to as a `diamond', moves as a solid
body. The head (blue) and tail (red) particles have a radius $R=1$
which defines the natural length scale.
The two side disks (green) have radius $a$ that is varied 
to alter the shape of the diamonds. The diamonds
are assumed to be governed by over damped dynamics. 
The equation of motion of disk $i$ within a diamond is
\begin{equation}
\frac{d{\bf r}_i}{dt}=V_0 {\bf \hat{n}}_i+\mu \sum_j {\bf F}_{ij},
\end{equation}
where ${\bf r}_i$ is the coordinate of the disk, $V_0$ is the speed due to the 
active force, ${\bf \hat{n}}_i$ is a unit vector from the
tail to the head, $\mu$ is the mobility parameter and ${\bf F}_{ij}$
is the radial harmonic interaction force 
$-k(|{\bf r}_i-{\bf r}_j|-R_i-R_j){\bf \hat{r}}_{ij}$ 
between pairs of disks. This force maintains a fixed separation
between disks within a diamond. 
Here $R_i$ and $R_j$ are the radii of the disks, $k$ is the spring constant
and ${\bf \hat{r}}_{ij}$ is a unit vector from disk $j$ to disk $i$.
In our simulations we have defined $V_0=1$, and used values of 
$\mu=1$, $k=100$ and a time step $\Delta t =10^{-3}$ time units.
Reducing the timestep did not strongly influence the observed behaviour.
We have also assumed that the mobility parameter $\mu$ that relates force to
velocity is the same for all disks, irrespective of their size or position
within the diamond.

The active
force that acts on each disk
results in a speed $V_0$ for the diamond as a whole.
 If a constant
torque, $\tau_r$, is applied to the diamond then the resulting constant 
angular velocity, $\omega$, is readily shown to be given by 
$\omega=\tau_r/2(R+a)^2$. Even though inertia is ignored in this over damped
limit, there is resistance to rotation that is shape dependent.

\begin{figure}
\resizebox{85.0mm}{!}{\includegraphics[clip]{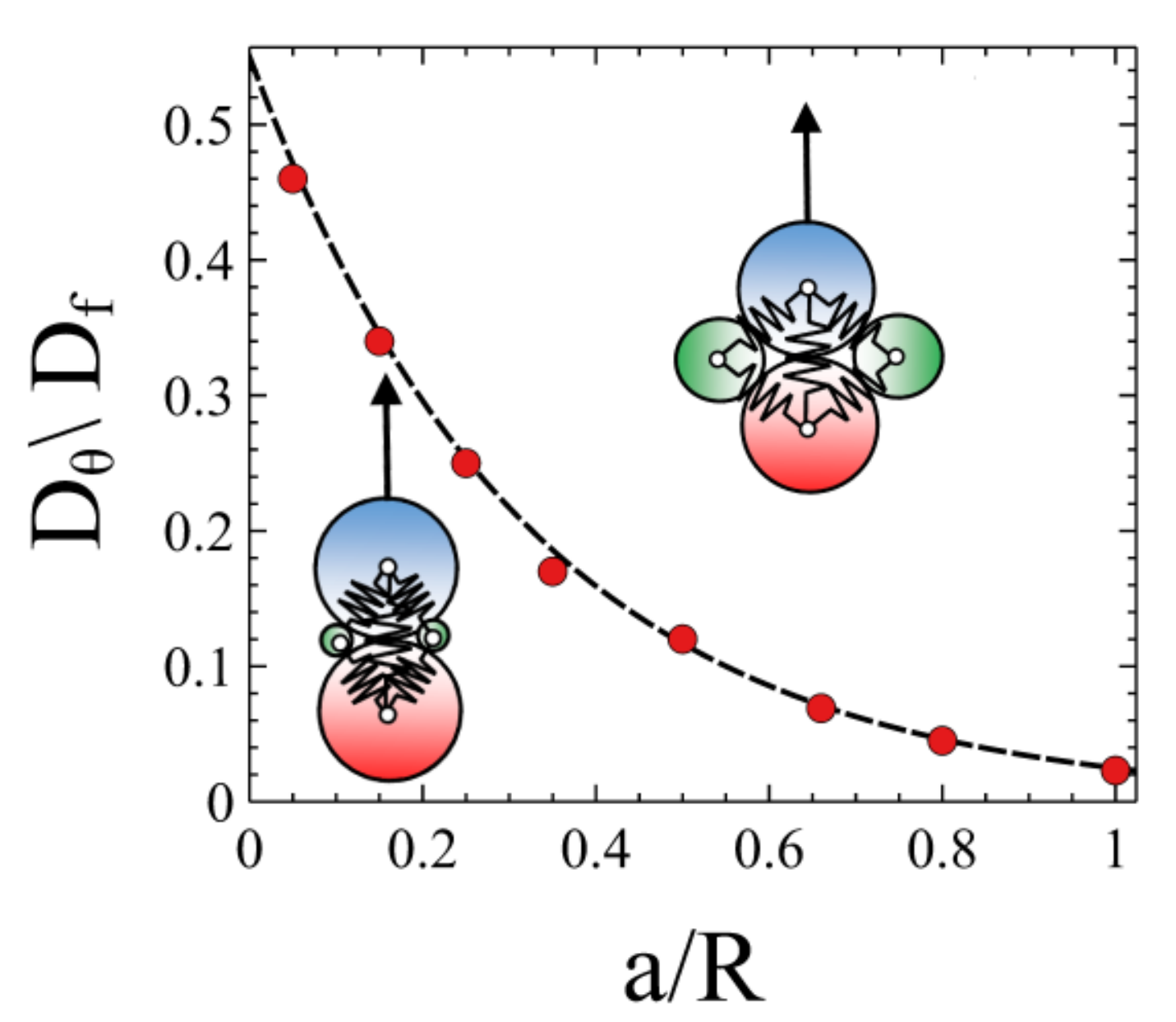}}
\caption{\label{fig1} 
The insets show the configuration of two composite particle `diamonds' with
different values of the shape parameter $a/R$, as indicated
by their positions on the $x$-axis. The springs
ensure that the diamonds move as solid objects.  An active force is added to each disk in the direction from the tail (red) to the head (blue). 
A random force is added to the
head and tail disks to induce angular diffusion.
The main panel shows the variation of the ratio of the
angular and linear diffusion coefficients as a function of $a/R$.}
\end{figure}

In order to introduce rotational diffusion, an additional random force is added in Eqn. 1 to the head and tail disks; each
force component $\eta_{\alpha}$ is assumed to be Gaussian with correlator 
$<\eta_{\alpha}(t)\eta_{\beta}(t')>= 2 D_f \delta_{\alpha,\beta}\delta(t-t')$, where 
$\alpha$ and $\beta$ are Cartesian coordinates and $D_f$ is a linear diffusion
constant that characterises the fluctuation of the applied
force. The resulting angular fluctuations can be measured and are
found to be diffusive with angular diffusion constant $D_\theta$.
Because of the shape dependence of the resistance to rotation, $D_\theta$
and $D_f$ are related to the ratio $a/R$ as shown in the main panel
of Fig 1. In what follows we define a linear Peclet number 
resulting from the applied random force $Pe^f=V_0 R/D_f$.
The corresponding angular Peclet number ${Pe}^{\theta}=V_0/RD_{\theta}$
can be obtained from the data shown in Fig. 1. In our simulations we
keep $V_0=1$ fixed and vary the noise strength to change the Peclet number.

\begin{figure}
\resizebox{85.0mm}{!}{\includegraphics[clip]{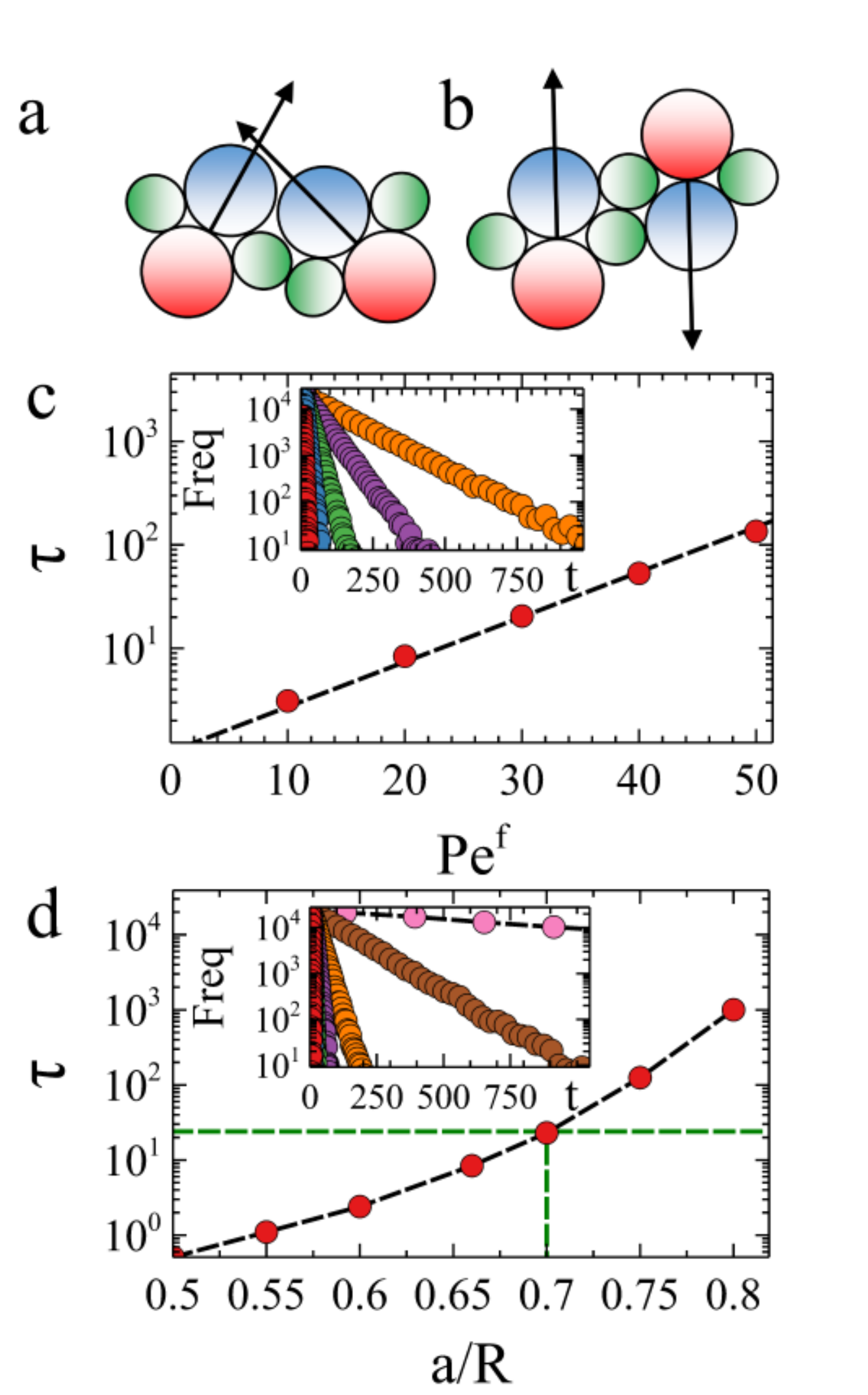}}
\caption{\label{fig2} 
The upper panel shows two possible configurations for a pair of diamonds: 
(a) glider configuration, (b) spinner configuration. The glider is stable
and propagates in the direction of the two head particles. The spinner
does not propagate but rotates on the spot. The survival time distributions 
for stable rotation
are shown in the insets to (c) and (d), for a range of Peclet numbers and $a/R$. The observed exponential 
decay can be characterised
by a lifetime $\tau$ shown in the main panels to (c) and (d). 
The stability of the spinners grows exponentially
with increasing Peclet number and grows even more strongly with increasing $a/R$.
The horizontal line in (d) shows when the lifetime of a spinner becomes comparable
with the time taken for a diamond to travel from one cluster to the next, as discussed
in the main text.}
\end{figure}

A single diamond behaves as an active Brownian particle with speed $V_0$
and angular diffusion $D_{\theta}$. If two diamonds collide,
they are assumed to repel due to repulsive springs which act radially between overlapping disks. The response to these forces is also treated in the over damped limit,
with the size-independent mobility parameter $\mu$ and the spring constant $k$ as above.

A distinctive feature of this study is that the diamonds, as a result of
their shape, can lock together exhibiting
collective dynamics of as few as two particles.
For example, consider the configurations shown in Fig. 2. The individual diamonds are constructed from disks with
a size ratio $a/R=2/3$. This is the closest shape that mimics a single disk;
namely, a circle of radius $2R$ just enclosing the diamond. In the left
hand configuration, Fig. 2a, the two diamonds lock together in such a way that
the pair move together at a constant speed but the angular fluctuations
are somewhat suppressed. Such an object travels around a periodic system indefinitely,
resembling a `glider' configuration in Conway's Game of Life \cite{Gardner}. The 
effective ${Pe}^{\theta}$ is much increased because of the suppression of
angular fluctuations due to the rotational resistance.

On the other
hand, the right hand configuration, Fig. 2b, allows the pair of particles to
remain in contact and rotate on the spot, suppressing translational
motion. Such a configuration is long lived due to the lack of inertia
in the over damped limit. These `spinners' eventually separate with a 
survival time that is given by an exponential distribution.
The lifetime is found to depend on the $Pe^f$, as shown in Fig. 2c. The
exponential growth of the lifetime with $Pe^f$ exhibits an Arrhenius behaviour
with the noise playing the role of an effective temperature. High $Pe^f$ spinners
thus remain stable in these configurations for longer resulting in a 
reduced effective Peclet number
compared with an isolated diamond. Figure 2d also shows how the lifetime of a spinner
depends on the shape parameter $a/R$. The dependence is even greater than exponential.
Given this strong influence of particle shape on few-body systems, we now ask
how shape effects influence the collective behaviour of many-particle
systems. 

The shape of the diamond is controlled by the ratio $a/R$.
For $a/R<1/4$ the side surface is concave; in the limit $a/R \to 0$ the
diamond resembles a dumbbell. For $a/R = 1/4$ the diamond has sides
that are relatively flat, mimicking a short rod. As noted above,
for $a/R=2/3$, the diamond corresponds as closely as possible to a circle,
whereas for $a/R>2/3$ the side-lobes start to dominate.
We have simulated multiple diamonds for a range of $a/R$. Each system has
2500 diamonds and the region has a size $L$ so that a $50\%$ by area filling
fraction is maintained. This definition ignores small variations due to
excluded area affects. Periodic boundary conditions are employed in both directions. 

\begin{figure}
\resizebox{85.0mm}{!}{\includegraphics[clip]{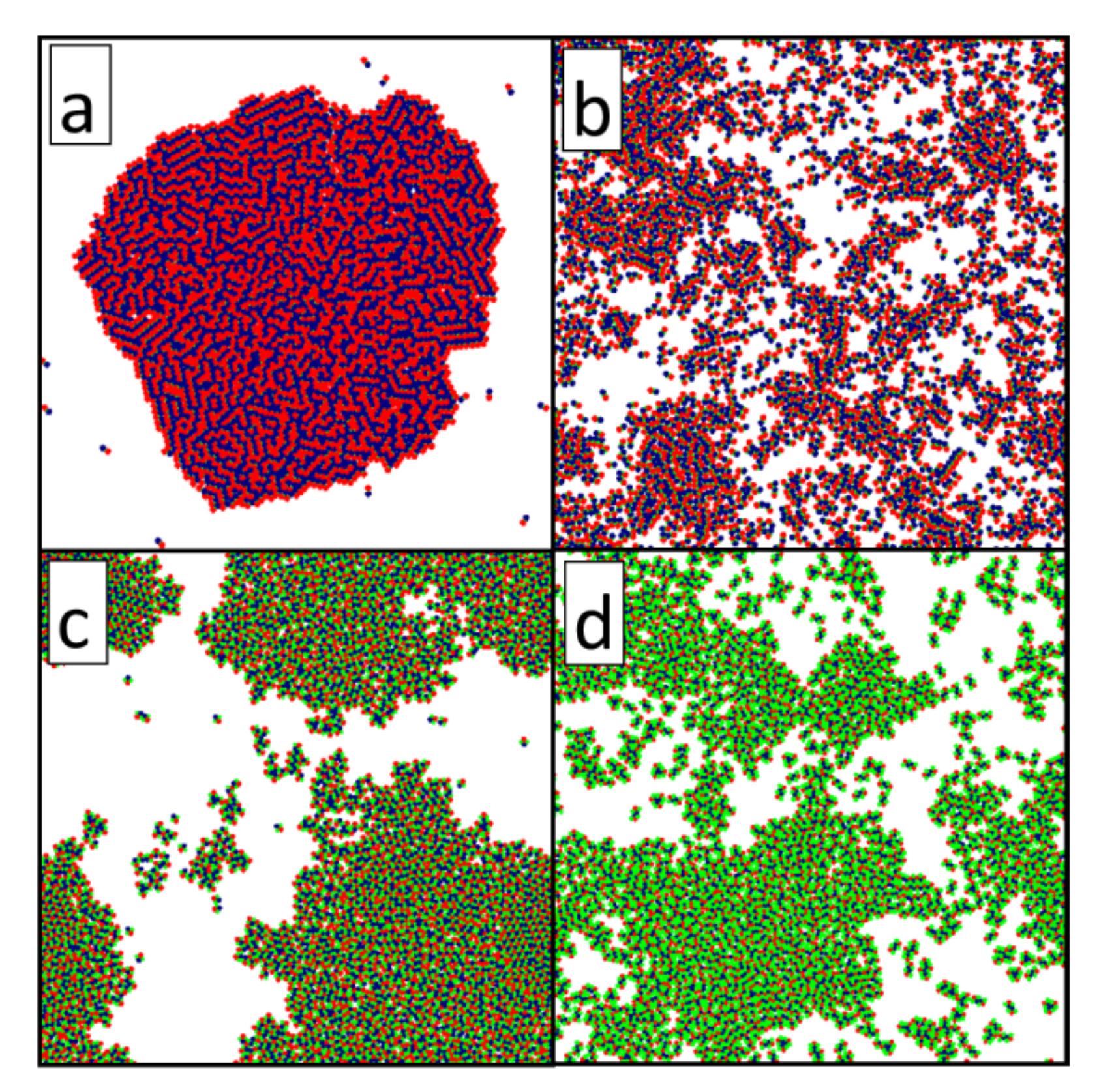}}
\caption{\label{fig3} 
Snapshots taken from simulations in the steady state for different
values of the shape parameter $a/R$: (a) 1/20 (b) 1/4 (c) 2/3 (d) 4/5.}
\end{figure}

Figure 3 shows a series of snapshots taken in the steady state for four different
values of $a/R$, all at $Pe^f=20$. Figure 3a is for $a/R=1/20$. Here a dense
crystalline cluster forms which slowly rotates about its centre. Figure 3b is for $a/R=1/4$. In this case there are only short-lived transient clusters and the
system remains homogeneous on the average over time. In Fig. 3c, $a/R=2/3$,
and a dynamic, fluid-like cluster persists with accompanying swirling
motion. Finally, Fig. 3d is for $a/R=4/5$ in which the tendency
to form large clusters appears less pronounced than for $a/R=2/3$. It is also striking that in between the clusters there are significant numbers
of spinners like the ones investigated in Fig 2. For movies
see the supplemental information \cite{Movies}.

In active systems with no attractive interactions, clustering
results from steric hindrance and a competition between the rate at which particles
arrive in and leave from a region of space (MIPS).
For simple disks, MIPS is controlled by the filling fraction and the
Peclet number. 

In order to quantify the observed behaviour in the diamond model, we have defined an order
parameter to measure density fluctuations. We calculate a course-grained 
density map based on the centre of each diamond and by dividing the
system into 400 squares. The system is allowed to relax for 9000 time units and
we then calculate the time-averaged
variance of these density fluctuations over the next 1000 time units. 
Such a measure is sensitive to
the overall structure but is not influenced by collective motion of
the clusters.

In the case of our non-circular particles, the ability of
a diamond to push past a neighbouring diamond or to leave a 
cluster depends on the particle geometry. In
addition, the tendency of a particle to propagate can be hindered
by local ordering as, for example, in the case of the spinners. Consequently, the
phase diagram has a rich structure.

\begin{figure}
\resizebox{85.0mm}{!}{\includegraphics[clip]{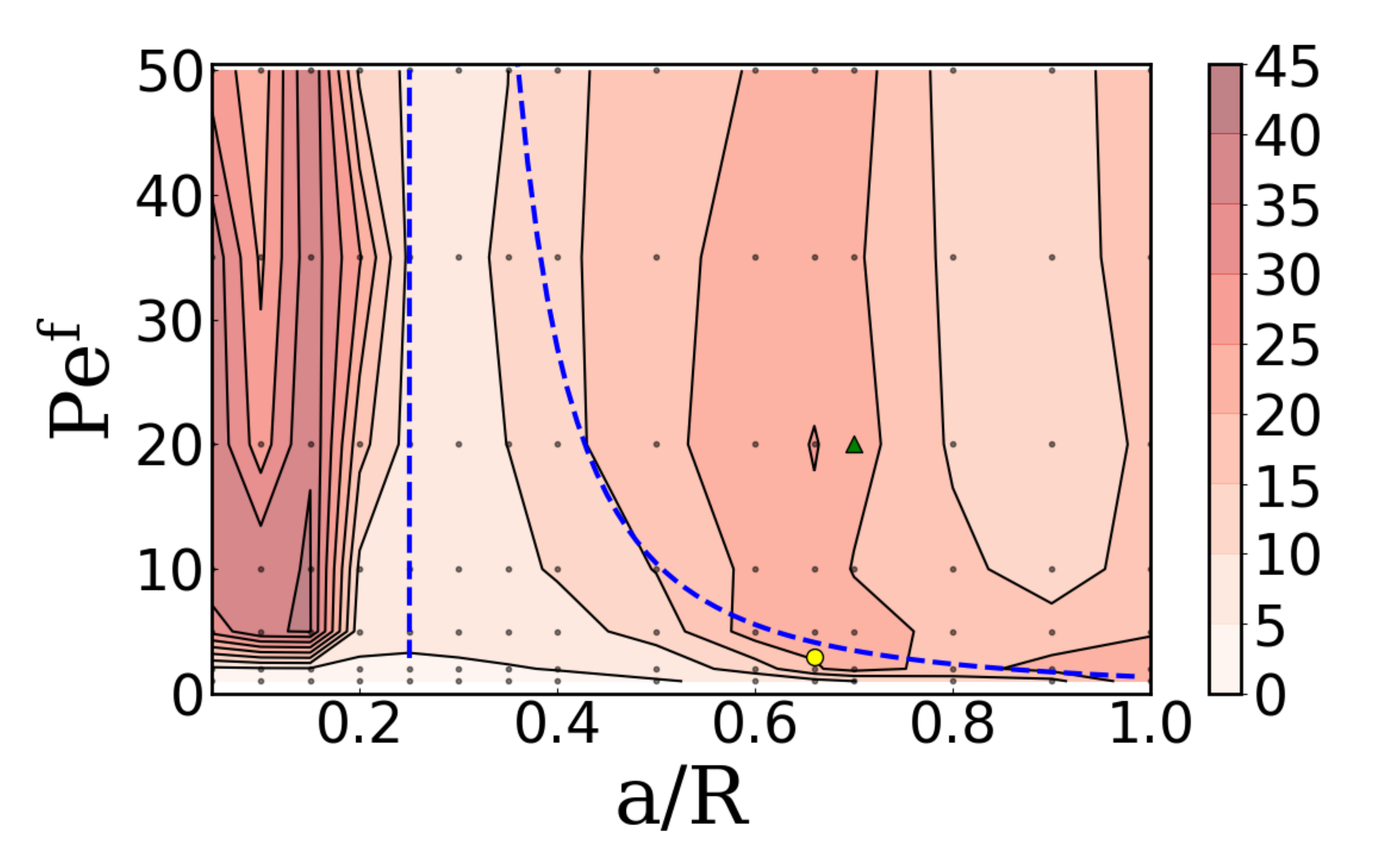}}
\caption{\label{fig4} 
Phase diagram showing the degree of clustering as a function of
Peclet number, $Pe^f$, and shape parameter, $a/R$. The colour scheme
goes from low (light) to high (dark) degrees of clustering, as measured
by the order parameter defined in the main text. The yellow circle represents
the expected onset of MIPS for circular disks. The green triangle shows where the spinner lifetime is approximately equal to the time taken to travel between
clusters at $Pe^f = 20$. The dotted blue lines represent expected transitions in behaviour based on geometrical arguments (see details in the main text). The black dots represent individual simulation runs.}
\end{figure}

Figure 4 shows the phase diagram in the space of ($a/R,Pe^f$) for a
fixed filling fraction of $50\%$.
At small $a/R$ the particles enter a crystallised arrangement which persists up to a $a/R$ of 0.2. 
A crystal represents the highest density packing of particles composed of 2 spheres. 
This high density suppresses fluctuations resulting in a very stable configuration that enhances MIPS by preventing the rearrangement of particles. 
Figure 3a illustrates this, where particles once joined to the main cluster remain attached. Such behaviour is similar to that observed in active 
dumbbells \cite{Cugliandolo}. 
The addition of small side particles does not disrupt the possibility of this configuration  provided $a/R \le 0.2$. 
This is because the side particles are not big enough to prevent a close packed arrangement of the larger particles.
There is therefore no strain applied to the crystal configuration. As the side particles increase slightly above $a/R = 0.2$ 
(the sides are still concave at this point) the large particles are forced slightly further apart. This creates a strain 
in the crystal, which decreases the stability of the packing. Fluctuations due to particle noise allow the relative position 
of particles to move more easily. 

The limiting case for this crystallization occurs when $a/R \sim 0.25$. 
At this point the edge of the small particle is aligned with the tangent to the two larger particles. 
As two particles come together they cannot interlock, simply sliding past one another. The phase diagram in 
Fig. 4 indicates that this coincides with a dramatic decrease in stable clustering. 
The left hand side of this diagram can therefore be understood in purely geometrical terms.

As $a/R \to 2/3$, the diamonds, as closely as possible
within this model, resemble disks and a transition analogous to MIPS for disks results.
The onset of clustering occurs at a value of the $Pe^{\theta}$ in close agreement
with that obtained in disk simulations \cite{Fily}. The yellow dot in Fig. 4 shows this point in terms of the corresponding $Pe^f$ obtained from Fig.1.
Around this value of $a/R$ there is an extended region in the
phase diagram of MIPS-like clustering. 
However, if circular particles are made even slightly ellipsoidal, it is known
that the MIPS mechanism breaks down \cite{Grossmann}.
This occurs because any non-circular shape experiences a torque about
its centre of mass that disrupts the polar boundary layer. In contrast,
our diamonds {\em can} maintain a polar boundary layer. We now demonstrate that
this difference arises due to the surface of a diamond having both
concave and convex regions, allowing interlocking of particles.

The boundary between rod-like (transient clusters) and disk-like (MIPS) 
behaviour can be estimated by
considering the time taken for one diamond to move past another. From
simple trigonometry, the distance a side-lobe protrudes beyond the tangent to
the two large disks (Fig. 1) is $\Delta=R|(2 \alpha + \alpha^2)^\frac{1}{2}+\alpha -1|$,
where $\alpha=a/R$.
Over a time $t$ the sideways motion is diffusive, so to travel a distance $\Delta$
requires a time given by the equation $\Delta^2 =2 D_f t$. 
This time can be compared with the one required for two diamonds to travel passed one another $(t=2R/V_0)$.
This comparison gives an estimate of the boundary between sliding and locking motion.
In terms of $Pe^f$ we find $Pe^f=4/((2 \alpha + \alpha^2)^\frac{1}{2}+\alpha -1)^2$.
This simple model, with no adjustable parameters, is shown by the right hand dotted blue line on Fig. 4, in good agreement
with the simulated data.

Increasing $a/R$ further results in a weakening of the tendency to cluster, this despite the apparent increase in stability of the 
two diamond configurations with increasing $a/R$. In Fig. 3d and the corresponding supplementary movies
one does not observe a significant number of the glider configuration (cf Fig. 2a). Though the more stable of the two diamond configurations,
any gliders that form quickly disintegrate as they collide with larger clusters. However, a large number of the spinning configurations, 
together with other small clusters of 3 or 4 are observed. Once a pair of diamonds
forms a spinner they no longer propagate, as shown in Fig. 2d, thus reducing the effective
Peclet number of the entire system. This reduction in number of free particles contributes to a decreased stability of the larger clusters.

To test this idea we estimate where the spinner survival time might start to become significant when compared to the time taken for
a diamond to propagate between clusters. To determine a typical 
distance between clusters, we assume that all particles occupy a circular region with filling fraction of 0.8 (for loosely packed disks). If the total area filling fraction is $\phi=1/2$
then the area of the cluster is $A_c =\phi L^2 / 0.8$ giving a radius
$R_c=(\phi L^2/ 0.8 \pi)^\frac{1}{2}$. The corresponding distance
between clusters is $L-2 Rc \approx 24$. From Fig. 2d this gives an $a/R$ value of about
0.7 as shown by the green triangle in Fig. 4, confirming the significance of localised
spinning clusters for the decrease in degree of clustering for higher values of $a/R$. 

Our simulations highlight the importance of shape in active matter systems.
The topology of our diamonds give rise to a range of novel collective
behaviours. It is shown that even two particles can result in spinning clusters
with lifetimes that exhibit an Arrhenius-like dependence on the Peclet number
and a strong dependence on the shape parameter $a/R$.
MIPS induced clustering is observed for a wide range of $a/R$ and can be predicted using simple scaling arguments.  
The fact that MIPS is suppressed for high $a/R$ also shows the role of 
few-body interactions; small clusters can form but they do not propagate, 
reducing the effective Peclet number. 
Convex and concave particle shapes may be particularly relevant in
soft active matter, where interparticle forces can result in complex
deformations. Phase behaviour in such systems
is likely to differ significantly from that of the simple
particle geometries more commonly studied. 

\begin{acknowledgements}
M.I.S. gratefully acknowledges a Royal Society University Research Fellowship.
\end{acknowledgements}

\end{document}